\documentstyle[prd,aps]{revtex}

\newcommand{\bea}{\begin{eqnarray}}
\newcommand{\eea}{\end{eqnarray}}

\begin{document}

\draft
\twocolumn[\hsize\textwidth\columnwidth\hsize\csname
@twocolumnfalse\endcsname

\title{On the ${1/ 3}$ factor in the CMB Sachs-Wolfe effect}
\author{J. Hwang${}^{(a,b)}$,
        T. Padmanabhan${}^{(c)}$,
        O. Lahav${}^{(b)}$ and
        H. Noh${}^{(d,b)}$
}
\address{
         ${}^{(a)}$ Department of Astronomy and Atmospheric Sciences,
                    Kyungpook National University, Taegu, Korea \\
         ${}^{(b)}$ Institute of Astronomy, Madingley Road, Cambridge, UK \\
         ${}^{(c)}$ Inter-University Centre for Astronomy and Astrophysics,
                    Post Bag 4, Ganeshkind, Pune 411 007, India \\
         ${}^{(d)}$ Korea Astronomy Observatory,
                    San 36-1, Whaam-dong, Yusung-gu, Daejon, Korea
}
\date{\today}
\maketitle

\begin{abstract}

We point out that a pseudo-Newtonian interpretations of the $(1/3)$ 
factor in the Sachs-Wolfe effect, which relates the fluctuations in 
temperature and potential, $(\delta T/ T) = (1/ 3) \delta \Phi$, is
not supported by the General Relativistic analysis. 
Dividing the full gravitational effect into separate parts depends on
the choice of time slicing (gauge) and there exist infinitely many 
different choices.
More importantly, interpreting the parts as being due to the gravitational
redshift and the time dilation is not justified in the rigorous
relativistic perturbation theory.
We suggest to regard the $(1/3)$ factor as the General Relativistic result 
which applies in a restricted situation of adiabatic perturbation in 
the $K = 0 = \Lambda$ model with the last scattering occuring in the 
matter dominated era. 
{}For an isocurvature initial condition the corresponding result, 
$(\delta T/ T) = 2 \delta \Phi$, has a different numerical coefficient.

\end{abstract}

\noindent
\pacs{PACS number(s): 98.70.Vc, 98.80.Hw}

\vskip2pc]

\section{Introduction}

Some of the exact results in General Relativity (GR) can be given
intuitive interpretations in terms of Newtonian concepts even
though the structures of these two theories are very different.
(i) The simplest example is the ``derivation'' of Schwarzschild radius
by equating the Newtonian escape velocity to the velocity of light;
one gets the correct numerical factor though the argument is not correct. 
(ii) The collapse time for a uniform density
dust sphere under self gravity computed by Newtonian theory turns
out to be exactly equal to the General Relativistic result.
(iii) The gravitational force on a material particle located inside
the empty region of a spherical shell of matter vanishes
in Newtonian theory.
This is usually explained by the fact that the force in Newtonian gravity
falls as $(1/r^2)$ while the amount of matter intercepted by a
a cone with fixed solid angle grows as $(r^2)$ thereby leading to
cancellation of forces due to opposite pairs of material in the shell
\cite{Binney-Tremaine-1987}.
Incredibly enough this result is true in GR as well though the
force law in GR is not strictly $(1/r^2)$ while the material
intercepted by a fixed solid angle does increase as $(r^2)$.
(iv) At a more subtle level, one can obtain the time-time component of
Einstein's equations for FLRW
(Friedmann-Lema\^itre-Robertson-Walker) universe from Newtonian energy
conservation argument {\it if} we take the potential energy of
a spherically symmetric mass distribution to be $-(GM/R)$.
The argument is again 
invalid in GR especially since Birkoff's theorem is applicable only 
for empty regions outside a mass distribution. 
In all the four cases mentioned above, these 
are accidental coincidences and the heuristic arguments 
have no deeper significance. 

Recently, another pseudo-Newtonian argument has surfaced to
explain the origin of $(1/3)$ factor in the Sachs-Wolfe effect in
cosmic microwave background (CMB) radiation \cite{Sachs-Wolfe-1967}.
The following heuristic argument is often seen in the literature
including several textbooks
\cite{Peacock-1991,Coles-Lucchin-1995,%
White-Hu-1997,%
Longair-1998,%
Bergstrom-Goobar-1999,%
Peacock-1999,%
Weinberg-2001}:
$(\delta T/T) = ({1/3}) \delta \Phi = \delta \Phi - ({2/3}) \delta \Phi$
where the first term arises from the gravitational redshift
whereas the second term comes from the time dilation of the
temperature perturbation
${\delta T/ T} = - {\delta a/ a} = - {(2/ 3)} ({\delta t/ t})
= - ({2/ 3}) \delta \Phi$ where $a$ is the scale factor.
{}For a similar argument see \cite{Peebles-1980}. 
We point out in this paper that the above decomposition is motivated
by a certain slicing condition (temporal gauge choice) out of
infinitely many possible choices.
More importantly, however, 
the interpretation of the terms as above cannot be
explained even in that slicing or in any other slicing
in the context of the correct relativistic analyses, see below eq. (\ref{ZSG}).
Hence, as the heuristic argument and interpretation are not supported
by rigorous theory, we believe such are spurious and have at best the
same status as the examples we mentioned in the first paragraph.

Before proceeding with a rigorous analysis of the situation,
it is worth pointing out that there are some fundamental difficulties
in providing {\it any} Newtonian interpretation of the Sachs-Wolfe effect.
Given any metric in GR, the Newtonian limit can be
rigorously established if the metric can be reduced to the form
$g_{00}= - (1 + 2\Phi)$ and $g_{\alpha\beta}=\delta_{\alpha\beta}$
in some coordinate system.
When one attempts to do this with FLRW metric, [see \cite{Padmanabhan-1996}],
one finds that the procedure is valid only for a region of size much 
smaller than the Hubble radius $d_H(t)\equiv (\dot a/a)^{-1}$.
This merely reiterates the fact that at scales bigger than the
Hubble radius one requires the full machinary of GR and 
- in particular - one needs to grapple with issues of gauge.
Since the Sachs-Wolfe effect arises from scales which are bigger than
Hubble radius at the epoch of recombination one can anticipate that
{\it any} Newtonian interpretation could have problem. 
This conclusion is also supported from the fact that one can consistently
choose a gauge in perturbed FLRW universe in which $g_{00} = -1$.
There is no Newtonian potential available in this gauge.
The key reason for the above pseudo-derivation to work is because
one introduced an ill-defined quantity $(\delta a/a)$ and
linked it with ${\delta T/ T}$ at one end [by $Ta=$ constant]
and with ${\delta t/ t}$ at the other end [by $a\propto t^{2/3}$].
There is no rigorous interpretation of this quantity
$(\delta a/a)$ possible in Newtonian gravity, or for that matter
in the gauge invariant treatments of FLRW perturbations. 
In handling the cosmological perturbation
one can adopt a heuristic argument using $\delta a$.
However, in this way, one often ends up with an equation which
is wrong (especially in a medium with 
pressure)\footnote{
        This is in contrast with the perturbed Hubble parameter $\delta H$ 
        which has a rigorous geometric and kinematic meanings, and in fact 
        is quite useful to handle the cosmological perturbation in a 
        heuristic looking but fully rigorous manner \cite{Hwang-Hyun-1994}.
        }. 
This is not surprising because perturbing the background system
necessarily loses some degrees of freedom compared with 
perturbing the full system.

\section{Temperature anisotropy in an arbitrary gauge}

We begin by obtaining the expression for the temperature anisotropy
without imposing {\it any} choice of gauge so that we can study
the results in the most general setting.
We consider a spatially homogeneous and isotropic
metric with the scalar-type perturbations
\bea
   & & ds^2 = - a^2 ( 1 + 2 \alpha ) d \eta^2
       - 2 a^2 \beta_{,\alpha} d \eta d x^\alpha
   \nonumber \\
   & & \qquad
       + a^2 \Big[ g^{(3)}_{\alpha\beta} ( 1 + 2 \varphi )
       + 2 \gamma_{,\alpha |\beta} \Big] d x^\alpha d x^\beta.
   \label{metric}
\eea
This represents a fairly general perturbed metric and no specific gauge
has been chosen.
The variables $\alpha$, $\beta$, $\varphi$ and $\gamma$ are spacetime
dependent scalar-type metric perturbations, and a vertical bar indicates
a covariant derivative based on the background three-space
comoving metric $g^{(3)}_{\alpha\beta}$.
We introduce the variable $\chi \equiv a (\beta + a \dot \gamma)$
which gives the shear of the normal-hypersurface.
$\varphi$ is proportional to the perturbed curvature of the hypersurface,
and $\alpha$ is the perturbed lapse function; we use notations in
\cite{Bardeen-1988,Hwang-Noh-1999-SW}.
The combination $\chi$ and the rest of the variables used
in the following are spatially gauge-invariant \cite{Bardeen-1988}.
The variables depend, however, on the temporal gauge (coordinate)
transformation which corresponds to choosing the spatial hypersurface,
i.e., the time slicing.
Thus we have the freedom to impose a temporal gauge (slicing) condition
which could be used as an advantage to handle the problems conveniently.
The prime and the overdot indicate the time derivative based on
$\eta$ and $t$, respectively, with $dt \equiv a d \eta$.

The most general expression of the Sachs-Wolfe effect from the
scalar-type perturbation can be found in eq. (15) of \cite{Hwang-Noh-1999-SW}.
Ignoring the Doppler effects due to the observer's motion, and the
emitting event (which is subdominant at large angular scales)
the observable temperature anisotropy becomes
\bea
   & & {\delta T \over T} \Big|_O
       = \Big( \alpha_\chi + {\delta T_\chi \over T} \Big) \Big|_E
       + \int^O_E \left( \alpha_\chi - \varphi_\chi \right)^\prime dy,
   \label{deltaT-3}
\eea
where the integration is along the photon's null geodesic path from
the emitted ($E$) epoch to the observed ($O$) epoch.
The variables $\alpha_\chi \equiv \alpha - \dot \chi$,
$\delta T_\chi \equiv \delta T + H T \chi$, and
$\varphi_\chi \equiv \varphi - H \chi$ are gauge-invariant
combinations\footnote{
       {}For the gauge transformation properties,
       see eq. (2) in \cite{Hwang-Noh-1999-SW}.}.
$\varphi_\chi$ becomes $\varphi$ in the zero-shear gauge (often called
the Newtonian gauge) which sets $\chi \equiv 0$, etc.
Using Bardeen's notation in \cite{Bardeen-1980} we have
$\alpha_\chi \equiv \Phi_A$ and $\varphi_\chi \equiv \Phi_H$,
\cite{SW-GI}.
The gauge invariance of a combination assures that there remains no
gauge (coordinate) mode and that its value remains the same in any gauge.
However, it does not guarantee a certian gauge-invariant variable has
an associated intrinsic physical meaning independent of the slicing condition.
{}For example, there exist several variables $\varphi_\chi$,
$\varphi_{\delta T}$ $\equiv \varphi + \delta T/T$,
$\varphi_v$ $\equiv \varphi - (aH/k) v$ where
$v$ is a velocity related variable and $k$ is a wave number,
etc., which are all gauge invariant.
(An exception is $\varphi_\alpha$ based on the synchronous gauge
which fixes $\alpha = 0$).
They all reduce to $\varphi$ under the corresponding gauge condition
which sets the variable in the subscript equal to zero.
Although all these gauge invariant variables
(we can make infinitely many different combinations)
are the curvature variables in different time slicing (temporal gauge)
we can safely regard them as completely different variables.

Equation (\ref{deltaT-3}) was derived from the geodesic
equations in the spacetime
of eq. (\ref{metric}) {\it without} fixing the gauge condition and
{\it without} using the gravitational field equations.
In the literature, the two terms in the RHS are often called
the Sachs-Wolfe effect and the integrated Sachs-Wolfe effect, respectively.
At this point it may be appropriate to quote a comment
in \cite{Sachs-Wolfe-1967} which looks
presient in the context of main point of this paper:
``{\sl We emphasize again that in a generic gravitational field
       one cannot distinguish gravitational redshifts from Doppler
       shifts by any standard recipe; thus our division of equation
       \dots has only a heuristic significance.}''

We shall now assume that:
(i) the anisotropic stress can be ignored, so that we have\footnote{
        See eq. (8) in \cite{Hwang-Noh-1999-SW}.
        }
$\alpha_\chi = - \varphi_\chi$;
(ii) $K = 0 = \Lambda$, and
(iii) the medium is an ideal fluid with constant $w \equiv p/\mu$
(where $p$ is the pressure and $\mu$ is the energy density)
so that the growing solution of $\varphi_\chi$ remains constant
in time\footnote{
       See eq. (18) in \cite{Hwang-Noh-1999-SW}.
       }.
Given {\it all} these assumptions the integrated Sachs-Wolfe term vanishes,
so that:
\bea
   & & {\delta T \over T} \Big|_O
       = \Big( - \varphi + 2 H \chi + {1 \over 4} \delta_{(\gamma)}
       \Big) \Big|_E,
   \label{gauge-ready-form}
\eea
where we used ${\delta T/ T} |_E = ({1/ 4} )\delta_{(\gamma)} |_E$
with $\delta_{(\gamma)}$ $\equiv \delta \mu_{(\gamma)}/\mu_{(\gamma)}$
denoting fractional energy-density perturbation
of the photons.
Now, the RHS is written without fixing the gauge yet,
thus in a sort of gauge-ready form, but the sum is gauge-invariant.
In this form we could understand why and how such a decomposition
into the intrinsic temperature perturbation, the gravitational redshift
(or the time dilation) is dependent on the temporal
slicing (gauge) condition of the spacetime.
Although $\varphi$ and $\chi$ have meaning as perturbed curvature
and shear of the normal three-hypersurface,
and $\delta_{(\gamma)}$ looks like a energy-density perturbation
of photons, these variables acquire such a meaning
{\it only after} fixing the temporal gauge (time slicing) condition
where we have infinite choices.
As mentioned before the same variable evaluated in a different slicing
(gauge) condition in general behaves as a completely different variable.

Equation (\ref{gauge-ready-form}) can be written in a suggestive
form as\footnote{
       With the wave number $k$ appearing in the equation the variables
       can be regarded as the Fourier transformed ones.
       To the linear order the same equations in configuration space
       remain valid in Fourier space as well.
       Thus, we ignore specific symbols distinguishing the variables
       in the two spaces.
       }
\bea
   & & {\delta T \over T} \Big|_O
       = \Big( - \varphi_\chi - {aH \over k} v_\chi
       + {1 \over 4} \delta_{(\gamma)v} \Big) \Big|_E,
   \label{form-1}
\eea
where $v_\chi \equiv v - (k/ a) \chi$
        and $\delta_{(\gamma)v} \equiv \delta_{(\gamma)} + 4 (aH/ k) v$.
        $\delta_{(\gamma)v}$ is the same as $\delta_{(\gamma)}$ in the
        comoving gauge which sets the velocity variable $v \equiv 0$;
        in a pressureless matter the test particles follow geodesics and hence
         the comoving gauge is equivalent to the synchronous gauge which fixes
        $\alpha \equiv 0$.
Thus, in this form the variables are viewed
(evaluated) in mixed slicing (gauge).
We can show that each of these gauge-invariant variables
most closely resembles the Newtonian counterparts as:
$-\varphi_\chi$, $\delta_v$, and $v_\chi$ most closely reproduce
the behavior of the perturbed gravitational
potential $\delta \Phi$, the perturbed density
$\delta_N \equiv {\delta \rho / \rho}$, and the perturbed velocity
$\delta v$ in the Newtonian context,
\cite{Harrison-Nariai,Bardeen-1980,Hwang-Noh-1999-Newtonian}.

It should be obvious from the above two equations and discussion
that the actual form of the terms in the right hand side depends
very much on the gauge.
It is best  not to yield to the temptation of interpreting the
individual terms ``physically'' - let alone try to fix the numerical
prefactors.
But if one insists on doing so, then the most natural choice is to
interpret the first term $- \varphi_\chi$ as due to gravitational redshift
(we have $- \varphi_\chi = \alpha_\chi = \delta \Phi$),
the second term as due to Doppler effect and the third
as arising from radiation field.
As we shall see in the next subsection,
even this interpretation is fraught with danger
but at least the coefficient of $- \varphi_\chi$ is now unity.

\section{Adiabatic perturbations}

We consider a system with radiation ($\gamma$) and matter ($m$).
The adiabatic condition
\bea
   & & S \equiv \delta_{(m)} - {3 \over 4} \delta_{(\gamma)} = 0,
\eea
implies
\bea
   & & \delta \equiv {\delta \mu \over \mu}
       = {1 + R \over 1 + 4R/3} \delta_{(\gamma)}, \quad
       R \equiv {3 \mu_{(m)} \over 4 \mu_{(\gamma)}},
\eea
thus
we have\footnote{See eq. (6) in \cite{Hwang-Noh-1999-SW}.
        }
$\delta_{(\gamma)v} \sim \delta_v
= {2 \over 3} ( {k/ aH} )^2 \varphi_\chi$
which is subdominant in large angular scales corresponding to
the large-scale ${k/ aH} \ll 1$ at $E$.
We have\footnote{
        See eq. (7) in \cite{Hwang-Noh-1999-SW}.
        }
\bea
   & & v_\chi = - {2 \over 3} {1 \over 1 + w} {k \over aH} \varphi_\chi,
   \label{v_chi-relation}
\eea
for the growing solution.
Thus, adding the first two terms in the RHS of eq. (\ref{form-1})
we finally have
\bea
   & & {\delta T^{({\rm Ad})} \over T} \Big|_O
       = - {1 + 3 w \over 3 (1 + w)} \varphi_\chi
       = - {1 \over 3} \varphi_\chi
       \Big|_E = {1 \over 3} \delta \Phi,
   \label{SW-adiabatic}
\eea
where the second equality follows by assuming matter domination at $E$,
so that 
$w = 0$\footnote{
       Using ${\delta a \over a} = {2 \over 3(1 + w)} {\delta t \over t}$
       the heuristic argument mentioned in the introduction also produces
       the result for general $w$ at the emission epoch $E$
       \cite{White-Hu-1997}.
       }.
In our case $\delta \Phi$ does not evolve in time.
We stress again that the result in eq. (\ref{SW-adiabatic})
is valid under many conditions mentioned above, especially the
ones above eq. (\ref{gauge-ready-form});
e.g., the simple result in eq. (\ref{SW-adiabatic}) does not hold
(for example) in a model with additional cosmological constant, and in such a case
we should go back to the general form in eq. (\ref{deltaT-3}).

It was often stressed in the literature that the $- (1/ 3)$ factor comes
directly from the metric part in the synchronous gauge
whereas it gets a contribution of $-1$ from the metric and $(2/ 3)$ from the
intrinsic temperature part in the zero-shear gauge
\cite{Bond-1996,Weinberg-2001}.
The origin of such gauge-dependent interpretations can be understood
simply by rewriting the RHS of eq. (\ref{gauge-ready-form})
or eq. (\ref{form-1}) in the respective gauge conditions.
In the zero-shear gauge, $\chi \equiv 0$, we have
\bea
   & & {\delta T \over T} \Big|_O
       = - \varphi_\chi + {1 \over 4} \delta_{(\gamma)\chi}.
   \label{ZSG}
\eea
Notice that in the large-scale the temperature part
$\delta_{(\gamma)\chi}$ is {\it dominated}
[when viewed in the comoving gauge, compare with eq. (\ref{form-1})]
by the metric, and does {\it not} behave like an ordinary temperature.
Instead, it gives ${2 \over 3} \varphi_\chi$, and we use
$-1 + {2 \over 3} = - {1 \over 3}$ to get the final result.
This is a rigorous argument, and one should not confuse this
with the heuristic one mentioned in the introduction;
except for the similar division into $-1$ and ${2 \over 3}$
the origins and the interpretations are completely different. 
Therefore, the heuristic interpretation is
not based on this zero-shear gauge analysis. 
The synchronous gauge coincides with the comoving gauge in 
the matter dominated era (MDE),
thus in the comoving gauge we have
\bea
   & & {\delta T \over T} \Big|_O
       = (- \varphi_v + 2 H \chi_v) + {1 \over 4} \delta_{(\gamma)v}.
   \label{SG}
\eea
{}For an ideal fluid with $w = {\rm constant}$ we have
$\varphi_v = {5 + 3 w \over 3 + 3 w} \varphi_\chi$\footnote{
        This follows from eq. (18) in \cite{Hwang-Noh-1999-SW}
        and the conservation property of $\varphi_v$ under an
        adiabatic condition: $\varphi_v = C$.
        Or, see eqs. (50,51) in \cite{Hwang-Noh-1999-Newtonian}.
        }
for the growing mode.
The temperature part now behaves like the conventionally known
temperature fluctuation, thus is negligible in the large angular scales
compared with the potential fluctuation.
Therefore, using eq. (\ref{v_chi-relation}) with
$v_\chi = - (k/a) \chi_v$
the metric part gives $-{1 \over 3} \varphi_\chi$ directly.
{}For the original derivation see \cite{Sachs-Wolfe-1967};
see also \cite{Padmanabhan-1993}.

\section{Isocurvature case}

The isocurvature condition is $\delta \mu_v \equiv 0$
under which we have $\varphi_\chi = 0$ and $\varphi_v = 0$.
This condition implies $S = - {3 \over 4} ( R^{-1} + 1 ) \delta_{(\gamma) v}$.
The isocurvature initial condition is imposed early
in the radiation dominated era (RDE).
Einstein's equations
give\footnote{This follows from eqs. (12,14,18)
              in \cite{Hwang-Noh-1999-Newtonian}.
             }
$\dot \varphi_v = - H (\mu + p)^{-1} \delta p_v$
which shows that the initial isocurvature
perturbation can generate $\varphi_v$.
{}For an isocurvature mode we have
$\delta p_v = - (1/ 3)\mu_{(m)} (1 + R)^{-1} S$.
Assuming the last scattering epoch $E$ occurred in the MDE
we have
\bea
   & & \varphi_v = {1 \over 3} S \int_{\rm RDE}^{\rm MDE}
       {dR \over ( R + 1 )^2} = {1 \over 3} S,
\eea
at $E$ where we used $S = {\rm constant}$ in the large-scale limit
\cite{Kodama-Sasaki-1986};
this argument was used by Liddle and Lyth in \cite{Liddle-Lyth-2000}.
Thus, in the MDE we have
$\varphi_\chi = {3 \over 5} \varphi_v = {1 \over 5} S$;
this shows the amount of curvature perturbation $\varphi_\chi$
in MDE {\it generated} from the initial isocurvature perturbation $S$ in RDE.
In MDE we have ${1 \over 4} \delta_{(\gamma)v} = - {1 \over 3} S$.
Therefore, from eq. (\ref{form-1}), using eq. (\ref{v_chi-relation})
and assuming MDE at $E$, we have
\bea
   & & {\delta T^{({\rm Iso})}\over T} \Big|_O
       = - {2 \over 5} S \Big|_E = - 2 \varphi_\chi \Big|_E = 2 \delta \Phi,
   \label{SW-isocurvature}
\eea
which is six times larger than the adiabatic result.
{}For original derivations, see below eq. (3.5) of
\cite{Efstathiou-Bond-1986} and below eq. (5.27) of \cite{Kodama-Sasaki-1987}.

\section{Discussions}

It is clear from  eq. (\ref{gauge-ready-form}),
that we can divide the terms in  different ways (which is
actually what the gauge choice is doing). Correspondingly there are
many different ways to reach {\it the same} final
results in eqs. (\ref{SW-adiabatic},\ref{SW-isocurvature}).
In other words, we have infinitely many different ways of
introducing slicing, thus viewing each variable in different gauges.
While doing an actual calculation we need to choose the gauge
(as we mentioned, a gauge-invariant variable is equivalent to
a variable based on a certain slicing condition which fixes the
gauge mode completely), but the final physical results should be 
the same independently of which gauge we have chosen.
Our results in eqs. (\ref{SW-adiabatic},\ref{SW-isocurvature})
are the final results where $\delta \Phi$ can be interpreted as
the perturbed Newtonian potential which is related to the density
contrast through Poisson's equation.

The pseudo-Newtonian method described in the introduction
is closely related to the decomposition in eq. (\ref{ZSG}).
However, as we have shown below eq. (\ref{ZSG})
such an interpretation is not supported by analyses in that gauge,
which is true even in the context of our gauge-ready form
in eq. (\ref{gauge-ready-form}). 
That is, interpreting the parts as being due to the gravitational
redshift and the time dilation is not justified in the rigorous
relativistic perturbation theory.
Hence it is difficult to imagine that such a heuristic argument 
captures the basic physics.
We believe it is yet another curious coincidence in General Relativity 
in which a pseudo-Newtonian argument does lead to the correct final result.  

In a classic book by Zel'dovich and Novikov \cite{Zeldovich-Novikov-1983}
we find a statement:
 ``{\sl However, \dots the gravitational shift contains the factor 1/3;
        it is still unclear how to interpret this coefficient classically.}''
which still seems to be true.

\subsection*{Acknowledgments}

We thank George Efstathiou and Kandaswamy Subramanian for useful discussions.
We also wish to thank Ed Bertschinger, Robert Brandenberger and Ruth Durrer 
for useful comments on this work. 
TP acknowledges hospitality of IoA, Cambridge.
HN was supported by grant No. 2000-0-113-001-3 from the
Basic Research Program of the KOSEF.
JH was supported by Grants KRF-2000-013-DA004, 2000-015-DP0080
and 2001-041-D00269.


\end{document}